\documentclass[aps,prd,twocolumn,groupedaddress,showpacs,tighten]{revtex4}
\usepackage{graphicx}
\usepackage{times}

\def\be{\begin{equation}}
\def\ee{\end{equation}}
\def\bea{\begin{eqnarray}}
\def\eea{\end{eqnarray}}
\def\ba{\begin{array}}
\def\ea{\end{array}}
\def\ghost#1{}
\def\dis{\displaystyle}

\def\beq{\begin{equation}}
\def\eeq{\end{equation}}
\def\bey{\begin{eqnarray}}
\def\eey{\end{eqnarray}}

\def\lsim{\mathrel{\raise.3ex\hbox{$<$\kern-.75em\lower1ex\hbox{$\sim$}}}}
\def\gsim{\mathrel{\raise.3ex\hbox{$>$\kern-.75em\lower1ex\hbox{$\sim$}}}}

\def\dis{\displaystyle}

\begin{document}

\title{Constraints on Light Dark Matter and \boldmath $\,U$ bosons, 
from \boldmath$\ \psi,\ \Upsilon, \ K^+,\ \pi^\circ,\ \eta\,$ and $\,\eta'\,$ decays}
\author{Pierre Fayet$^{1}$}

\affiliation{$^1$Laboratoire de Physique Th\'eorique de l'ENS, UMR 8549 CNRS, 24 rue Lhomond, 75231 Paris Cedex 05, France}
\date{July 28, 2006}

\begin{abstract}
{
Following searches for photinos and very light gravitinos 
in invisible decays of $\psi$ and $\Upsilon$, 
we discuss new limits on Light Dark Matter and $\,U$ bosons,
from $\,\psi\,$ and $\,\Upsilon\,$ decays,
as well as rare decays of $\,K^+\,$ and invisible decays of $\,\pi^\circ,\ \eta\,$ and $\,\eta'$ ... .
The new limits involving the {\it vector} couplings of the $U$ to quarks turn out, 
not surprisingly, to be much less restrictive than existing ones on {\it axial} 
couplings, from an axionlike behavior of a light $U$ boson, 
tested in $\,\psi\to\gamma \,U,\ \Upsilon\to \gamma \,U$ and $K^+\to\pi^+ U$
decays \,(or as compared to the limit from parity-violation in atomic physics,
in the presence of an axial coupling to the electron).
Altogether the hypothesis of light $U$ bosons, and light dark matter particles,
remains compatible with particle physics constraints, 
while allowing for the appropriate annihilation cross sections required, both at freeze-out \,(for the relic abundance)\,  and nowadays
\,(if $e^+$ from LDM annihilations are at the origin of the 511 keV line from the galactic bulge).
}
\end{abstract}

\pacs{\ 12.60.Cn, 13.66.Hk, 13.20.-v,
13.20.Cz, 13.20.Eb, 13.20.Gd,
14.70.Pw, 
95.35.+d  \vspace{2mm}\\
LPTENS-06/31}
\maketitle

\vspace{-5mm}

Theories beyond the standard model generally include a number of new 
particles, such as the neutralinos of the supersymmetric standard model \cite{susy,grav},
the lightest of which, stable by virtue of $R$-parity,
is now a leading dark matter candidate.
Or new neutral bosons, such as the spin-0 axion \cite{axion},
or a new neutral spin-1 gauge boson $\,U$ which could be light and very weakly coupled \cite{fayet:1980rr}. This one can also play an essential role in the annihilations 
of light dark matter particles into $e^+e^-$ \cite{boehmfayet,fermion}, that may be at the origin at the 511 keV line from the galactic bulge \cite{integral,boehm511}.

\vspace{2mm}

A privileged way to search for such particles, especially in the case of supersymmetry, is to look for a ``missing energy'' 
signal, i.e. missing energy-mom\-en\-tum that would be carried away, in particular, 
by unobserved photinos or neutralinos, and gravitinos. Or also axions, $U$ bosons, cosmions,
or light dark matter particles, ... .
We shall discuss here some limits, in addition to those of 
\cite{fayet:1980rr, boehmfayet,fermion,pvat,cras} -- that the decays of the $\psi$ or the $\Upsilon$, or of the $K^+\!,\ \pi^\circ,\ \eta$ or $\eta'$ mesons, can impose on light dark matter (LDM) particles and $U$ bosons, and more specifically on the $U$ couplings to quarks and LDM particles.

\section{\vspace{1mm}
Limits on gravitino and photino production
\hbox{from invisible decays 
of \boldmath $\,\psi\,$ and $\,\Upsilon$}}

The decays of quarkonium states such as the $\psi\,$ and the $\,\Upsilon$, 
\,and $\,e^+e^-$ annihilations, were used very early to search for 
``invisible'' particles, and constrain their properties.
Limits on invisible decay modes of the $\psi$ and the $\Upsilon$ have been known 
in fact for a long time.

\vspace{2mm}

As discussed in \cite{psi}, a search for the invisible decays of the $\psi$, 
identified from the observation, in $\psi'\,$ decays, of a $\pi^+$ and $\pi^-$ with a well-defined invariant mass, according to
\be
\ba{l}
\psi' \ \,\to\,\   \pi^+ \pi^-\ \psi \ \ \ \ ,
\\
\hspace{23mm} \hookrightarrow\  {\rm invisible}
\ea
\ee
led to the upper limit \cite{breid} 
\be
\label{psiinv}
B\,(\psi \to {\rm invisible}) \ <\  7 \ 10^{-3} \ \ ,
\ee
to be compared with $\,B\,(\psi \to e^+e^-)\simeq \,7\pm 1 \,\%$ \,(taken to be 
$>6\,\%$, \,in fact its present value).  This can be used to 
constrain the associated production of (ultralight) spin-$\frac{3}{2}$ gravitinos and (light) spin-$\frac{1}{2}$ photinos
from the expression of the {\it gravitational\,} decay rate
\be
\Gamma(\psi\to{\rm gravitino} + {\rm photino})\ \,\propto \ \,\frac{G_{\rm Newton}\,\alpha}{m_{3/2}^2}\ \ ,
\ee
supersymmetry being then spontaneously broken ``at a low scale''.

\vspace{2mm}
In such a situation, only the ``longitudinal'' $\,\pm\frac{1}{2}\,$ polarisation states 
of the massive but very light gravitino are actually taking part in the decay, bringing in a factor $\sqrt{2/3}\ \ k^\mu/m_{3/2}$ from the gravitino wave function
\cite{grav}.
More precisely, a very light gravitino may be viewed has having with the photon and 
its superpartner the photino  \,a non-diagonal $q^2$-dependent charge-like effective coupling 
\be
\label{eff}
e_{\rm eff}(q^2)\ =\ 
\frac{\kappa\ q^2}{m_{3/2}\,\sqrt 6}\ =\ \frac{q^2}{d}\ \ ,
\ee
where $\,\kappa=(8\pi\,G_{\rm Newton})^{1/2}\simeq 4. 1\ 10^{-19}\ \hbox{GeV}^{-1}$, and $d$ is the supersymmetry-breaking scale parameter \footnote{One now often replaces $d$ by $F$ such that $d^2/2=F^2$, then defining 
the susy-breaking scale as
$\,\Lambda_{\rm ss} =\sqrt F = \sqrt d/2^{1/4}$.}.
This $q^2$ factor compensates the $1/q^2$ from the photon propagator, 
leading to an effective local 4-fermion interaction with charged particles proportional to 
$\,\kappa\,e/(m_{3/2}\,\sqrt 6)=e/d$ (see \cite{psi} for details).
The same amplitude can also be found, equivalently, by considering the production of a massless spin-1/2 goldstino 
in the corresponding spontaneously-broken globally supersymmetric theory, 
in agreement with the equivalence theorem of supersymmetry \cite{grav}.

\vspace{1mm}

Altogether one gets the simple relation \footnote{The distinction 
between photino and antiphotino, gravitino and antigravitino, which made sense for very light particles, is no longer appropriate in the presence of sizeable neutralino Majorana masses.}
\be
\label{rate}
\frac{B(\psi\to{\rm gravitino} + {\rm photino})}{B(\psi\to e^+e^-)}\ \simeq\ \frac{e_{\rm eff}^{\,2}(m_\psi^{\,2})}{e^2}\ \ ,
\ee
which led to the first direct experimental lower limit on the gravitino mass, 
$m_{3/2} > 1.5\ 10^{-8}$ eV. Or equivalently on the supersymmetry-breaking scale, 
supersymmetry being broken ``at a low scale''.
This analysis and the resulting bounds, also reported in \cite{PDG}, although not really constraining yet, are at the starting point of the phenomenology of a very light gravitino, its mass fixing the effective strength of its interactions, 
and therefore the time at which it decouples from equilibrium, in the evolution of the Universe.

\vspace{2mm}

A similar search for  
\be
\ba{l}
\Upsilon(nS) \ \ \to\ \  \pi^+ \pi^-\  \Upsilon(1S)
\\
\hspace{35mm} \hookrightarrow\ {\rm invisible}
\ea 
\ee
was performed a few years later by CLEO \cite{CLEO1}, leading to 
\be
B\,(\Upsilon(1S)\ \to\ \hbox{invisible})\ < \ 5\ \%\ \ ,
\ee
from $\Upsilon(2S)$ \,(or 8\,\% from $\Upsilon(3S)$).
This limit seems worse than (\ref{psiinv}) by a factor $\,\simeq 7$, 
\,further increased to $\,\simeq 18$ as it should be divided by the rate for $\Upsilon\to e^+e^-$, \,of only $\,\simeq 2.4\,\%$ (instead of $\,6\,\%$ for $\psi$). 
This is, however, more than compensated by an increased sensitivity to gravitino production, 
enhanced by  $(m_\Upsilon/m_\psi)^4\simeq 87$, owing to eqs.~(\ref{eff},\ref{rate}). 
Altogether the effective sensitivity is increased 
by almost 5, resulting in a lower limit 
$m_{3/2} >\, 3\ 10^{-8}$ eV, \,twice the one obtained from $\psi$.
\vspace{2mm}

This illustrates the interest of working 
both with precision and at high energies, to take advantage of the $\,q^2$
factor in $\,e_{\rm eff}(q^2)$, as we are effectively looking for an almost local dimension-6\, four-fermion interaction, whose effects increase with energy.

\vspace{1mm}
This was  further pursued by searches in $e^+e^-$ annihilations at higher energies
(PEP and PETRA), for \cite{photinoprod,photinoprodexp}
\be
\label{eeino}
\left\{\ 
\ba{ccl}
e^+e^-&\to& \ \hbox{gravitino\ +\ photino}\ \ ,
\vspace{1mm}\\
e^+e^-&\to& \ \ \hbox{photino\ +\ photino}\ \ .
\ea \right.
\ee
These reactions may be signed by the associated emission of a (soft) single photon, or 
of a photon produced in the decay of a massive photino into photon + gravitino.
This raised the lower limit on $\,m_{3/2}\,$ up to about $10^{-5}$ eV,
corresponding to a supersymmetry-breaking scale $\sqrt d> 240$ GeV
(for a not-too-heavy photino and under conditions precised in \cite{PDG,photinoprod}).

\vspace{1mm}
Reactions such as (\ref{eeino}) now represent
``searches for the production of dark matter particles at accelerators''. 
The inverse reactions describe the annihilations of two dark matter candidates, 
here neutralinos, into $e^+e^-$  or other particles, 
ultimately responsible for the relic abundance of dark matter, in a way which depends on the mass spectrum of the various particles involved in the annihilation.

\vspace{1mm}
It is not our purpose to discuss more gravitinos or neutralinos, but 
how experiments searching for missing energy (or $\,\gamma$ + missing energy)
in $q\bar q$ or $e^+e^-$ 
annihilations may be used to constrain other ``invisible'' neutral particles such as ``cosmions'' (neutral particles of a few GeV's, with rather strong
couplings to hadrons), as discussed in \cite{cosmion}, with interactions and properties prefigurating to some extent those of light dark matter particles.
Or, more interestingly for us, light dark matter (LDM) particles \cite{boehmfayet, fermion}.
They require new powerful annihilation mechanisms responsible for large annihilation cross-sections at freeze-out (typically $\,<\!\sigma_{\rm ann}\,v_{\rm rel}/c\!>\ \approx\,$ a few pb), as necessary to get the appropriate relic abundance corresponding to $\,\Omega_{\rm dm}\simeq 22\,\%$. 
The new neutral light \hbox{spin-1} gauge boson $U$, very weakly coupled at least to ordinary particles \cite{fayet:1980rr}, 
may then lead to the required annihilation cross sections,
significantly larger than weak-interaction cross sections, at these energies.
Conversely, the same mechanisms could also be responsible for the pair-production 
of LDM particles at accelerators, in particle interactions or decays involving missing
energy (or photon + missing energy) in the final state.
(Invisible quarkonium decays as already used to look for supersymmetric particles
and cosmions \cite{psi,breid,PDG,CLEO1,cosmion} were reconsidered recently in \cite{elrath}.)

\section{\vspace{1mm}
Production of \boldmath $\,U\,$ bosons 
\vspace{1mm}
\hbox{and light dark matter particles} \hbox{in \boldmath  $\,\psi\,$ and \boldmath $\,\Upsilon\,$ decays}}

As the $U$ boson we consider, mediating
$\,q\bar q\,$ and $\,e^+e^-\,$ annihilation processes, is supposed to be light, there
is in general no interest, when searching for the pair-production of 
light dark matter particles, in trying to work systematically at higher energies, 
in contrast with searches for gravitinos and photinos, or more generally neutralinos
\footnote{There is an exception here
for a light $U$ with axial couplings, as in the low mass regime the wave function of a longitudinally-polarised $U$ can  include a large factor $\simeq\,\frac{k^\mu}{m_U}$.
It would be then be produced very much as a spin-0 particle having pseudoscalar couplings with quarks and leptons
$f_{q,l\ p} = \frac{2\,m_{q,l}}{m_U}\ f_{q,l\ A}$ \cite{fayet:1980rr,fermion,cras}.}.
These processes may be naturally compared with electromagnetic ones,
an upper bound on the production of invisible neutrals in $q \bar q$ or $e^+e^-$ 
annihilations being translated into a bound 
$\,|c_\chi \,f_{q,e}|\ <\ ...\ e^2$, \ $c_\chi$ and $f$ denoting the $U$ couplings to dark matter particles, and quarks or electrons, respectively.

\vspace{2mm}

As the $\psi$ and $\Upsilon$ are \hbox{spin-1} $\,q\bar q\,$ states with $C=-$,
their direct decays through the virtual production of a single $U$ boson,
such as
\be
\psi\ \ (\hbox{or}\ \Upsilon)\ \ \to\ \ \hbox{invisible}\ \ \ \chi\,\chi\ \ , \ \ ...\ 
\ee
i.e. decays into invisible particles only,
can only occur through the {\it vector} coupling of the $U$ to a $c$ or $b$ quark, $f_{qV}$. This will lead to upper limits on the products $\,|c_\chi\,f_{qV}|$, 
\,to be discussed later.

\vspace{2mm}

In between, we note that
possible {\it axial} couplings $f_{qA}$ would contribute to the radiative decays 
of the $\psi$ and the $\Upsilon$, e.g.
\be
\left\{\ 
\ba{ccl}
\psi\ \ (\hbox{or}\ \Upsilon)\ &\to&\ \ \ \gamma\ U\ \ , 
\vspace{1mm}\\
\psi\ \ (\hbox{or}\ \Upsilon)\ &\to&\ \gamma\ \chi\,\chi\ \ , \ \ ...\ \ \ .
\ea \right.
\ee
We shall consider first the direct production of a real $U$ boson,
through its axial couplings $\,f_{qA}$, in these radiative decays of $\psi$ and $\,\Upsilon$. 
This leads, as we already know, to very strong constraints on the 
$\,f_{qA}$'s, so that for the time being we can postpone a discussion 
of $\gamma\chi\chi$ production, which would constrain $\,|c_\chi\,f_{qA}|$.

\subsection{Radiative production of $U$ bosons (through {axial} couplings)}

Two mechanisms, already discussed in 1980, are here essential to understand the rates at which a light $U$ boson could be produced in the radiative decays of the $\,\psi$ or the $\,\Upsilon$ \cite{fayet:1980rr}.

\vspace{2mm}

{\bf 1)}\  At first, the wave fonction of a {\it longitudinally-polarised light spin-1 gauge boson $U$} includes a large factor $\epsilon^\mu \simeq k^\mu/m_U$. 
A light spin-1 $U$ boson then
behaves very much as a spin-0 (quasi-Goldstone) boson
\footnote{Just as a very light spin-$\frac{3}{2}$ gravitino behaves very much as a spin-$\frac{1}{2}$ goldstino.}, 
having 
{\it effective pseudoscalar couplings} with quarks and leptons
\be
\label{pseudo}
f_{q,l\ \,p}\ =\ \frac{2\,m_{q,l}}{m_U}\ \ f_{q,l\ \,A}\ \ .
\ee
As a result, if the $SU(2)\times U(1)\times$ extra-$U(1)$ gauge symmetry is spontaneously broken to 
$\,U(1)_{\rm QED}$ 
through the v.e.v.'s of two electroweak Higgs doublets $h_1$ and $h_2$ only, as in 
the simplest supersymmetric theories, 
these effective pseudoscalar couplings would be the same as those of a standard axion \cite{axion}. A situation that became, some time later, excluded by experimental results \cite{edwards,crystal}.

\vspace{2mm}
But\, {\bf 2)}\  in the presence of an extra Higgs singlet $\phi$ transforming 
under the extra-$U(1)$, that would acquire a (possibly large) v.e.v.,
the additional $U(1)$ symmetry may be broken at a scale larger than the electroweak scale, possibly even ``at a large scale'', for $<\!\phi\!>$\, (much) larger than $\,v_F/\sqrt 2\simeq$ 174 GeV. 
The rates for directly producing a light $U$, or its equivalent spin-0 particle, would then be smaller (or possibly very small).
The spin-1 $U$ would behave in that case like a doublet-singlet combination
\be
\left\{
\ba{l}
\ \ \ \ \ \ \cos\eta \ \ \hbox{(old standard-axion-like pseudoscalar)}
\vspace{2mm}\\
+\ \sin\eta \ \
\hbox{(new electroweak singlet $\phi$, uncoupled to $\ q,\, l$)}\ \, ,
\ea \right.
\ee
which corresponds precisely \cite{fayet:1980rr} to the mechanism by which the standard axion 
could be replaced by a new axion, called later ``invisible''.
Previous decay rates for producing a $U$ boson, 
instead of being the same as for a standard axion 
(or light $A$ in the MSSM language), were then  multiplied by a factor
\be
\cos^2\eta\ =\ r^2\ \leq\ 1\ \ .
\ee

The effective pseudoscalar couplings (\ref{pseudo}) to quarks and leptons
may be written as
\be
f_{q,l\ \,p}\ =\ 2^{1/4}\ G_F^{1/2}\ m_{q,l}\ \ (\,x\ \hbox{or\ $\Large \frac{1}{x}$}\,)\ \ r\ \ ,
\ee
and identified with those of a (non-standard) axion,
with the following expression of the axial couplings of the $U$ to quarks and leptons,
\be
\ba{ccc}
f_{q,l\ \,A}\ &=&\ \dis 2^{-3/4}\ G_F^{1/2}\ m_U\ \ (\,x\ \hbox{or\ $\Large \frac{1}{x}$}\,)\ \ r\ \ ,
\vspace{2mm}\\
&\simeq&\ \dis 2\ 10^{-6}\ \ m_U(\hbox{MeV})\ \ (\,x\ \hbox{or\ $\Large \frac{1}{x}$}\,)\ \ r\ \ .
\ea
\ee

\vspace{2mm}

The resulting production rates of $U$ bosons
in radiative decays of the $\psi$ and $\Upsilon$, computed as for an axion 
\cite{axion} from the ratios $\,\psi\to\gamma U\,/\,\psi\to e^+e^-$, \ or $\,\Upsilon\to\gamma U\,/\,\Upsilon\to e^+e^-$, read \cite{fayet:1980rr,cras,fermion}
\be
\left\{   \  \begin{array}{ccll}
B \ (\ \psi \ \to \ \gamma \, U\ ) &\simeq &
  \ 5 \ \ 10^{-5} \ \ \,\ r^2 \ x^2 \  \ \ &C_{\psi} \ \ , \nonumber 
\vspace{2mm}\cr
B\  (\ \Upsilon \ \to \ \gamma\,  U\ ) &\simeq &
 \  2 \ \ 10^{-4} \ \ \,(r^2/x^2) \ \  &C_{\Upsilon}   \ \ ,         \cr
\end{array}   \right.
\ee
$C_{\psi}$ and $C_{\Upsilon}$, expected to be larger than $1/2$, taking
into account QCD radiative and relativistic corrections. 
A $\,U\,$ boson decaying  into LDM particles 
(or $\,\nu\, \bar \nu\,$ pairs) would remain undetected.
From the experimental limits \cite{edwards,crystal,CLEO2}
\begin{equation}   \left\{   \  \begin{array}{ccl}
B\ (\ \psi \ \to \  \gamma \,+\,  \hbox{invisible}\ ) \ & <&\ \, 1.4 \ \ 10^{-5}\ \ ,  
\vspace{2mm}\cr
B\ (\ \Upsilon \ \to \ \gamma \,+\, \hbox{invisible}\ ) \  & < & \ \, 1.5 \ \ 10^{-5}
\ \ , \cr
\end{array}   \right. 
\end{equation}
we deduced 
$\ rx  < .75$ and $\ r/x <.4 $. \,This requires $r\lsim\frac{1}{2}$, i.e. that the 
additional $U(1)$ symmetry should in this case be broken at a scale $\,F$ 
at least of the order of twice the electroweak scale. 
These limits may be turned into upper limits on the axial couplings of a $U$ to $c$ and $b$ quarks, i.e.
\be
\label{limfcba}
f_{c A}\ <\ 1.5\ \,10^{-6}\ m_U\hbox{(MeV)}\ ,\ \ 
f_{b A}\ <\ 0.8\ \,10^{-6}\ m_U\hbox{(MeV)}\ ,
\ee
respectively. These axial couplings are constrained to be rather small, 
in a way which may be remembered approximately as
\be
\frac{f_{q A}^{\,2}}{m_U^{\,2}}\ \ \lsim\ \,\frac{G_F}{10}\ \ .
\ee

This discussion should of course be adapted, as considered elsewhere, 
if the $U$ decays preferentially into $e^+e^-$ instead of remaining invisible.
This could happen for $m_U<\,2\,m_\chi$,  with $U$ couplings to neutrinos small
as compared to electrons.

\vspace{2mm}

We also recall that the vector couplings $\,f_{qV}$ cannot contribute to the decays $\,\psi\to\gamma\,U,\ \Upsilon\to\gamma\,U$, 
and are not directly constrained in this way.

\subsection{Production of LDM particles 
\hbox{in \boldmath  $\,\psi\,$ and \boldmath $\,\Upsilon\,$ decays}}

We now return to the production of LDM particles in decays 
of the $\psi$ and the $\,\Upsilon$,
\be
\psi\ \ (\hbox{or})\ \ \Upsilon\ \ \to\ \ \hbox{invisible LDM particle pair}\ \ .
\ee

\vspace{2mm}
In addition to the $U$-exchange amplitudes considered throughout this paper, 
spin-0 dark matter particles $\,\varphi$ could also interact with quarks through an effective dimension-5 interaction \cite{boehmfayet,fermion} proportional to
\be
\label{phiq}
\frac{c_l\,c_r}{m_{F_q}}\ \ \varphi^*\varphi\ \ \overline {q_R}\,  q_L\ +\ \hbox{h. c.}\ \ .
\ee
$m_{{F_q}}$ denotes the mass of the heavy (e.g. mirror) quark whose 
exchanges may be responsible for the annihilation $q \bar q \to$ \hbox{spin-0} 
LDM particle pair
\,-- just as the similar exchange of a heavy electron $F_e$ could induce ($S$-wave) annihilations of spin-0 LDM particles into $e^+e^-$.
Such an interaction, if present at a significant level, 
could also contribute to invisible \,(or $\gamma$ + invisible) decay modes of $\Upsilon$ or other $q\bar q$ states, in addition to the $U$-exchange contributions.

\vspace{2mm}
We do not expect the ($C$-even) operators $\,q\bar q\,$ or $\,\bar q\gamma_5 q\,$  in 
(\ref{phiq}) to contribute to the invisible decays of $\,1^{--}$ $q\bar q\,$ states 
such as $\,\psi$ and $\Upsilon$. Spin-0 $\,\varphi\bar \varphi\,$ pairs could however be produced through (\ref{phiq}) in their radiative decays. 
There is here again an advantage in working at higher energies,
as the effects of the dimension-5 operator (\ref{phiq}), as compared to those induced by 
a light $U$ boson, grow with energy.
In particular, from the limit \cite{CLEO2}
\be
\label{upsilonrad}
B(\Upsilon\to\gamma+\hbox{invisible})\ <\ 3\ 10^{-5}\ \ ,
\ee
in which ``invisible'' means here a $\,\varphi\bar\varphi$ (or $\chi\chi$ or $\chi\bar\chi$ ) pair, we can deduce
a limit such as $\,|c_l c_r\ m_\Upsilon/m_{F_q}|< ...$\ , \,however not expected to be very constraining, as the rate for $\Upsilon\to \gamma\,\pi^+\pi^-$, for example,
is only about $6\ 10^{-5}$.
This means also, especially in view of the dimension-5 character of the spin-0
operator (\ref{phiq}),
returning to more sensitive higher-energy reactions
$\,e^+e^- \to\gamma\, +$ invisible, as considered 
in \cite{photinoprod,photinoprodexp,cosmion,boehmfayet} for supersymmetric particles, cosmions or  LDM particles; in view of constraining, this time, $|c_l c_r/m_{F_e}|$.

\vspace{2mm}
We now return to $U$ exchanges, the main object of our interest, mediating through their {\it vector} couplings to quarks the decay of a $\psi$ or $\,\Upsilon$ into an
invisible pair of LDM particles.
The virtual $U$ from $\,q\bar q\,$ annihilation may 
convert into a pair of cosmions, or spin-1/2 (Majorana or Dirac,
say $\chi\chi$ or $\chi\bar\chi$), or of spin-0 ($\varphi\bar\varphi$) LDM particles.
The decay amplitude for
\be
\psi\ (\hbox{or}\ \Upsilon) \ \ \to\ \ \varphi\bar\varphi \ \ ,
\ee
proportional to $f_{qV}$ times the $U$-charge $c_\varphi$, is {\it $C$-conserving};
the final state has $C=(-)^L = -$ with $L=1$. 
The decay
\be
\psi\ (\hbox{or}\ \Upsilon) \ \to \ \chi\chi\ \  (\hbox{or}\ \chi\bar\chi)\ \ 
\ \ ,
\ee
proportional to $f_{qV}$ times the axial coupling $\,c_\chi$ of the $U$ 
to the fermion (Majorana or Dirac) $\chi$, is {\it $C$-violating};
the final state has $\,C=(-)^{L+S} = +$ with $\,J=1$  and therefore $L=S=1$.
In both cases of spin-0 and axially-coupled spin-$\frac{1}{2}$ we have a {\it $P$-wave production of light dark matter particles} in the final state. This reflects
(exchanging the roles of initial and final states)
that such light dark matter particles undergo {\it $P$-wave} (rather than $S$-wave) {\it annihilations\,} into lighter $f\bar f$ pairs through a vector coupling $f_{qV}$
of the $U$, as discussed in \cite{fermion}.

\vspace{2mm}
On the other hand for vectorially-coupled dark matter,
decays $\,\psi\,$ (or $\Upsilon$) $\,\to \chi\bar\chi$,
which would be proportional to  $f_{qV}$ times a vector coupling $\,c_{\chi V}$ of the $U$ 
to a Dirac LDM fermion $\chi$, would be $C$-conserving. The final state has then $C=(-)^{L+S} = -$ with $\,J=1$, and therefore $L=0$ (or 2) with $S=1$, \,or $L=1,\,S=0$, allowing for 
$S$-wave production of LDM particles.
Conversely such a vector coupling of the $U$ to spin-$\frac{1}{2}$ Dirac LDM particles 
would allow for $S$-wave dark matter annihilations, a situation that we should normally 
avoid at least for LDM annihilations into $e^+e^-$  at freeze-out \cite{boehmfayet,fermion}, unless this $S$-wave contribution is 
kept small enough, so that $P$-wave annihilation be dominant at freeze-out.
($S$-wave contributions, however, could still play a role in today's annihilations of LDM particles
into $e^+e^-$ within the galactic bulge \cite{rasera}, possibly at the origin of the 511 keV $\gamma$-ray line observed by INTEGRAL \cite{integral}.)

\vspace{2mm}
We can now express the rates for $\,\varphi\bar\varphi$, $\chi\chi$ or $\chi\bar\chi$ production in $\psi$ decays as follows \footnote{For a Majorana spinor $\chi$, $c_\chi$ is defined as the $U$-charge of the chiral fermion $\chi_L$ associated with $\chi\,$, 
the axial coupling of the $U$ being written as 
$\ \frac{c_\chi}{2}\ \,\bar\chi\gamma^\mu\gamma_5\chi \ U_\mu\,$.}:
\be
\!\!\left\{\
\ba{lccl}
\dis
\hbox{spin-0:} &\dis
\frac{B(\psi\to \varphi\bar\varphi)}{B(\psi\to e^+e^-)} \!&\simeq&\! \dis
\frac{f_{cV}^{\,2}\,c_\varphi^{\,2}}{(\frac{2}{3}\ e^2)^2}\ \ \ \frac{1}{4}\ \ \ [\,\beta^3\,] \ ,\vspace{1mm}\\
\dis
\hbox{Maj., axial:} &\dis
\frac{B(\psi\to \chi\chi)}{B(\psi\to e^+e^-)} \!&\simeq&\! \dis
\frac{f_{cV}^{\,2}\,c_\chi^{\,2}}{(\frac{2}{3}\ e^2)^2}\ \ \ \frac{1}{2}\ \ \ [\,\beta^3\,] \ , \vspace{1mm}\\
\dis
\hbox{Dirac, axial:} &\dis
\frac{B(\psi\to \chi\bar \chi)}{B(\psi\to e^+e^-)} \!&\simeq&\! \dis
\frac{f_{cV}^{\,2}\,c_{\chi A}^{\,2}}{(\frac{2}{3}\ e^2)^2}\ \ \ \ \ [\,\beta^3\,] \ , \vspace{1mm}\\
\dis
\hbox{Dirac, vector:} &\dis
\frac{B(\psi\to \chi\bar\chi)}{B(\psi\to e^+e^-)}\!&\simeq&\! \dis
\frac{f_{cV}^{\,2}\,c_{\chi V}^{\,2}}{(\frac{2}{3}\ e^2)^2}\ \ \ [\,\frac{3 \beta-\beta^3}{2}\,]\,,
\ea \right.
\ee 
with $\ \beta=v_f/c\,=(1-4\,m_{(\varphi \ {\rm or}\ \chi)}^{\ \ 2}/m_\psi^{\,2})^{1/2}
\simeq 1$, as the dark matter particles considered are light compared to $\,m_\psi$.
Having this ratio smaller than $7\,10^{-3}/6\,10^{-2}\simeq .12\,$ 
\cite{psi,breid} requires 
\be
\label{limcfc}
|c_\chi\,f_{cV}|\ \lsim\ \left\{\ \ba{rl}
\ 4\ 10^{-2} &\ \hbox{(spin-0)}\ ,\vspace{1mm}\\
\ 3\ 10^{-2} &\ \hbox{(Majorana)}\ ,\vspace{1mm}\\
\ 2\ 10^{-2} & \ \hbox{(Dirac)}\ 
\ea \right.
\ee
(the latter limit being $\frac{2}{3}\ e^2 \times \sqrt{.12}$\ ).
\vspace{2mm}

The corresponding limits from $\Upsilon\,$, \,now governed by $\frac{1}{3}\ e^2 \times (5\,\%\,/\,2.4\,\%)^{1/2}\simeq 4.5\ 10^{-2}$, \,are weaker than those from the $\psi$, by a factor slightly larger than 2:
\be
\label{limcfb}
|c_\chi\,f_{bV}|\ \lsim\ \left\{\ \ba{rl}
\  9\ 10^{-2} &\ \hbox{(spin-0)}\ ,\vspace{1mm}\\
\ 6\ 10^{-2} &\ \hbox{(Majorana)}\ ,\vspace{1mm}\\
\ 4.5\ 10^{-2} & \ \hbox{(Dirac)}\ .
\ea \right.
\ee
This in contrast with the gravitino limit which was improved by a factor $\simeq 2.2$ 
by going from $\,\psi$ to $\Upsilon$, as it benefited from the $m_\Upsilon^{\,2}/m_\psi^{\,2}\simeq 9.3\,$ enhancement factor
in the amplitude, a factor no longer present for the amplitudes induced by the exchanges of a light $U$.
\vspace{2mm}

We shall in general also demand that $\,c_\chi<\sqrt{4\pi}\,$, 
\,so that the theory remains perturbative. In the rather extreme case 
for which $\,c_\chi$ would be taken as large as $\sqrt{4\pi}$, 
the above limit (\ref{limcfc}) would imply, for a Majorana $\,\chi$,
\be
\label{limfcv}
|f_{cV}|\ <\ .9\ 10^{-2}\ \ ,\ \ \hbox{i.e.}\ \ \frac{f_{cV}^{\,2}}{4\,\pi}\ <\ 6\ 10^{-6}\ \ .
\ee

\vspace{2mm}
Even in this case, is this really very constraining\,?
To get a feeling we recall the constraint from parity-violation effects in atomic physics
\cite{pvat}
\be
|f_{eA}\,f_{qV}|\ <\ (1.5\ \ \hbox{to}\ \ 3)\ \ 10^{-14}\ m_U(\hbox{MeV})^2\ \ ,
\ee
which expresses that
\be
\frac{|f_{eA}\,f_{qV}|}{m_U^{\,2}}\ \,<\ \,\frac{G_F}{300}\ \ .
\ee
In the presence of an axial coupling to the electron,
even as small as $\,\simeq 10^{-7}\ m_U$(MeV), 
this implies a very strict upper limit on the quark vectorial coupling,
\be
|f_{qV}|\ \lsim \ 3\ 10^{-7}\ m_U(\hbox{MeV})\ \ ,
\ee
corresponding to
\be
\frac{f_{qV}^{\,2}}{4\pi}\ \lsim\ 10^{-14}\ \ m_U(\hbox{MeV})^2\ \ .
\ee
\vspace{1mm}

In view of this, and of the other constraining limits (\ref{limfcba}) on the axial couplings to the $c$ and $b$ quarks, we may be tempted to stick to theories in which the $U$ has only vector couplings to quarks and leptons, no axial ones at all, 
as in a class of models discussed in \cite{vector}.

\vspace{2mm}
We can then compare the above limits (\ref{limcfc},\ref{limcfb},\ref{limfcv}) to the 
one on a vectorial coupling to the muon, derived from $\,g_\mu-2\,$ 
(assuming no special cancellation effect) namely, for $m_U<m_\mu$ \cite{fermion},
\be
\label{limfmuv}
|f_{\mu V}|\ <\ (.7\ \,\hbox{to}\ 1.5)\ 10^{-3}\ \ ,\ \ \hbox{i.e.}\ \ \frac{f_{\mu V}^{\,2}}{4\,\pi}\ <\ (.4\ \,\hbox{to}\ 1.8)\ 10^{-7}\  \ ,
\ee
The limits (\ref{limcfc},\ref{limcfb},\ref{limfcv}) appear less constraining than (\ref{limfmuv})
\,-- although they concern different quantities --\,
and are {\em a fortiori} not significantly restrictive, when we confront them 
to the constraint
\be
\label{sizecf}
|c_\chi\ f_e|\ \simeq 10^{-6}\ 
\frac{|m_U^{\,2}-4\,m_\chi^{\,2}|}{m_\chi\ (1.8\ {\rm MeV})}\ \ 
\left(B_{\rm ann}^{ee}\right)^{\frac{1}{2}}\,,
\ee
on 
$\,c_\chi\,f_e$ necessary  to get the large annihilation cross-sections 
$\chi\chi\to e^+e^-$ required at freeze time
\cite{fermion}. This is even more easily realised for a relatively light $U$ boson.
To give just an example, for $m_U= 10$ MeV and $m_\chi\simeq 4$ (or 6) MeV, 
$f_e$  would have to be $\gsim 10^{-6}$, depending on $\,c_\chi$ assumed to be 
$< \sqrt{4\pi}$. In other terms we can obtain the right annihilation cross sections with
an $f_e$ as small as $\approx 10^{-6}$, i.e. $f_e^{\,2}/(4\pi)$ as small as about $\approx 10^{-13}$, much smaller than the $\,\gsim 10^{-5}$ we are dealing with in
(\ref{limcfc},\ref{limcfb},\ref{limfcv}).

\vspace{2mm}

While interesting, these limits  (\ref{limcfc},\ref{limcfb},\ref{limfcv}) on the quark vector couplings $f_{qV}$,
and subsequent ones to be discussed soon from
$\,K^+,\ \pi^\circ,\ \eta$ or $\,\eta'$ decays, are, not surprisingly, much less restrictive than those constraining the axial quark couplings
\cite{fayet:1980rr,fermion}, or from parity-violation effects, in the presence of an axial coupling to the electron \cite{pvat}.

\section{\vspace{1mm}
Production of \boldmath $\,U\,$ bosons 
\vspace{1mm}
\hbox{and light dark matter particles} 
\vspace{1mm}
\hbox{in \boldmath  $\ \,K^+\to \pi^+\ +$ ``invisible'' \ \ decays}}

\subsection{Production of $U$ bosons through their axial couplings}

Let us now consider the possible production of ``invisible particles'' in $K^+$ decays, namely
\be
K^+\ \to\ \pi^+\ U\ \ ,
\ee
in which the $U$ could stay invisible as decaying into two LDM particles
if $m_U>2m_\chi$, or decay into $\,e^+e^-$ or $\,\nu\bar\nu\,$ pairs
(a $U$ decaying into $e^+e^-$ would lead to a new source of $K^+\to \pi^+e^+e^-$
events)
\cite{fayet:1980rr, fermion}.
The contribution of an axial coupling of the $U$ is here essential, 
especially as in the small mass regime a longitudinally-polarised spin-1
$U$ boson behaves very much 
as a spin-0 pseudoscalar having effective couplings with quarks and leptons as 
given by (\ref{pseudo}).

\vspace{2mm}

The strong experimental limit on the branching ratio \cite{kpiu}
\be
B(K^+\to \pi^+\,+\,\hbox{invisible}\,\ U)\ <\ (.73\ \,\hbox{to}\  \approx 1)\ 10^{-10} \ \ ,
\ee
for $\,m_U<100$ MeV, may then imply (depending on how this branching ratio
is evaluated) the quite restrictive upper limit
\be
\label{limfsa}
f_{s A}\ \lsim\ 2\ 10^{-7}\ m_U(\hbox{MeV})\ \ .
\ee
This corresponds to demanding that the effective pseudoscalar coupling to the $s$ quark 
(\ref{pseudo}),
\be
f_{s\,p}\ =\ \frac{2\,m_s}{m_U}\ \ f_{sA}\ \ ,
\ee
verify
\be
f_{s\,p}\ \lsim\ 6\ 10^{-5} \ \ ,\ \ \ \hbox{or}\ \ \ 
\frac{f_{s\,p}^{\,2}}{4\pi}\ \lsim\ 3\ 10^{-10}\ \ .
\ee
Even if one can discuss more what the precise value of the limit should be, 
this quark axial coupling is in any case quite strongly constrained.

\subsection{Production of $U$ bosons through their vector couplings}

Let us now come to the {\it vector} couplings, much less constrained especially 
for the smaller values of $m_U$. The decay amplitude for $K^+\to \pi^+$ 
+ massless spin-1 particle vanishes exactly. The amplitude for 
$K^+\to \pi^+ U$ through a vector coupling of a light $U$ is expected to
vanish proportionally to $m_U$. For larger values of $m_U$ (typically $\gsim m_\pi$), we can compare the decay rate to the 3-body one,
\be
B(K^+\to\pi^+  e^+e^-)\ = \  (2.88\pm 0.13)\ 10^{-7}\ \ ,
\ee
under the simplifying assumption that the latter proceeds mainly 
through the virtual production of a photon converting into $e^+e^-$.
Replacing this virtual photon (coupled to a $u$ or $c$ quark of charge $2e/3$, as we consider mainly penguin graphs), by a virtual $U$ vectorially coupled to this quark
(with coupling $f_{uV}=f_{cV}$),
we expect  
\be
\frac{B(K^+\to \pi^+ U)}{B(K^+\to \pi^+ e^+e^-)}\ \approx\ 
\frac{f_{uV}^{\,2}/(4\pi)}{\frac{1}{\pi}\ (\frac{2}{3}\, \alpha)^2}\ \approx\ 
\frac{9\,f_{uV}^{\,2}}{16\,\alpha^2}
\ \  .
\ee
i.e $\,B(K^+\to \pi^+ U)\ \approx \ 3\ 10^{-3}\ f_{qV}^{\,2}\,$.
(For smaller values of $m_U$ this decay rate should vanish proportionally to $m_U^{\,2}$.)

\vspace{2mm}

This should be compared with an experimental upper limit of the order of
\,(1 to a few) $10^{-9}$, for $m_U$ between about 170 and 240 MeV, leading in this mass interval to an upper bound
$\,|f_{qV}|\lsim 10^{-3}$, \,or $\,f_{qV}^{\,2}/(4\pi) \lsim 10^{-7}$.
For $\,m_U\simeq m_{\pi^\circ}$,
the limit on $\,K^+\to \pi^+\,+$ invisible $U$ would be the same as for $K^+\to \pi^+ +$ invisible $\pi^\circ$ \cite{artamonov}, 
namely about 21\,\% $\times \,2.7\ 10^{-7}\simeq 6\ 10^{-8}$, 
leading to $\,|f_{qV}|<5\ 10^{-3}$, or $\,f_{qV}^{\,2}/(4\pi)\, < \,1.5\  10^{-6}$.

\subsection{Production of light dark matter particles}

We now consider
\be
K^+\ \ \to \ \ \pi^+\ \chi\chi\ \ \ \ (\hbox{or}\ \ \varphi\bar\varphi,\ \ \hbox{or}\ \ \chi\bar\chi)\ \ ,
\ee
induced by the virtual production of a $U$ boson converting into $e^+e^-$.
We compare again to $K^+\ \to \pi^+e^+e^-$
(remembering that the amplitude for producing a real photon in $K^+\to\pi^+ \gamma$ vanishes).
Replacing this virtual photon by a virtual $U$ vectorially coupled to a $u$ or $c$ quark,
we can write, at least for the lighter LDM (and $U$) masses,
\be
\frac{B(K^+\to \pi^+ \chi\chi)}{B(K^+\to \pi^+ e^+e^-)}\ \approx\ \frac{1}{2}\ 
\frac{c_\chi^2f_{uV}^{\,2}/(4\pi)^2}{(\frac{2}{3}\,\alpha)^2}\ \approx\ \frac{1}{2}\ 
\frac{9\,c_\chi^2 f_{uV}^{\,2}}{4\,e^4}
\ \  .
\ee
The factor $\frac{1}{2}$, here associated with the pair-production of Majorana particles,  would be replaced by 1 for Dirac particles ($\chi\bar\chi$), and $\frac{1}{4}\,$ for 
spin-0 particles ($\varphi\bar\varphi$).

\vspace{2mm}
$K^+\to \pi^+ \nu\bar\nu$ has been measured with a branching ratio 
1.47{\scriptsize$ \ba{c} +1.30\\ -0.89\ea $}
$\times 10^{-10}\,$ \cite{kpiu}.
One gets, at the 90 \,\% c.l., the upper bound 
\be
B(K^+\to\pi^+ + \chi\chi \ \hbox{(or other invisible)})\ < \ 3.84\ \,10^{-10}\ \ ,
\ee
to be compared with
\be
B(K^+\to\pi^+ \, e^+e^-)\ = \  (2.88\pm 0.13)\ 10^{-7}\ \ .
\ee
This leads to a ratio invisible\,/\,$e^+e^-\,$ typically smaller 
than about $\,1.3\ 10^{-3}$.
Provided $m_\chi\,$ remains relatively small as compared to $m_\pi$
(as expected), and $\,m_U \lsim m_\pi$ (or in any case is not too heavy),
we can deduce the upper limit on the product $\,c_\chi\,f_{uV}$,
\be
\label{limcfuv}
|c_\chi\,f_{uV}|\ \lsim \ 3.5\ 10^{-2} \ e^2\ \simeq\ 3\ 10^{-3}\ \ ,
\ee
implying, in the extreme case for which  $|c_\chi|$ 
would be taken as large as $\sqrt{4\pi}$,
\be
\label{limfuv}
|f_{uV}|\ < 10^{-3} \ \ ,\ \ \ \ \hbox{or}\ \ \frac{f_{uV}^{\,2}}{4\pi}\ <\ 10^{-7}\ \ .
\ee

\vspace{2mm}

The very strict upper limit of a few $10^{-10}$ 
on $\,K^+\to\pi^+\nu\bar\nu$ only implies limits which are not very restrictive, in comparison with (\ref{limfcba},\ref{limfsa}) and (\ref{sizecf}).
This is thus perfectly compatible with the values of the 
coupling product $|c_\chi\,f_e|$ necessary to provide sufficiently large annihilation cross sections 
for light dark matter particles.

\section{\vspace{1mm}
Production of \boldmath $\,U\,$ bosons 
\vspace{1mm}
\hbox{and \,LDM\, particles\, 
in \boldmath $\,\pi^\circ,\ \eta\,$ or \boldmath $\,\eta'\,$ decays}}

\subsection{\boldmath $\,\pi^\circ\to UU\,$}

Let us now give briefly here a few other results from
$\pi^\circ,\ \eta$ and $\eta'$ decays.
The $\,\pi^\circ$ is an isospin-1 state which may be described as
$( u\bar u - d \bar d)/\sqrt 2$.
The decay rate for $\,\pi^\circ\to UU\,$ may be related
to $\,\pi^\circ \to \gamma\gamma$, taking into account both vector and axial couplings of the $U$ to $u$ and $d$ quarks. As vector currents have $C=-$ and axial ones $C=+$, there are no 
$V A$ interference terms in the amplitude, while $A A$ contributions are
expressed similarly to the $V V$ ones. 
Adding $V V$ and $A A$ contributions in the amplitude we have,
in the limit of small $m_U$ compared to $\,m_{\pi^\circ}/2\ $,
\be
\ba{ccl}
\displaystyle
\frac{B(\pi^\circ\to \,U \,U)}{B(\pi^\circ \to \,\gamma\gamma)}\ &\simeq&\
\displaystyle \left(\ \frac{f_{u}^2-f_{d}^2}{(2e/3)^2-(-e/3)^2}\ \right)^2
\vspace{4mm}\\
&\simeq& \
\displaystyle
\frac{9\ (f_{u}^2-f_{d}^2)^2}{e^4}\ \ .
\ea
\ee
denoting for simplicity  $\,f_u^2= f_{uV}^{\,2}+f_{uA}^{\,2},\ f_d^{\,2}= f_{dV}^{\,2} +f_{dA}^{\,2}$.

\vspace{2mm}

This may be compared with the experimental limit \cite{artamonov}
\be
B(\pi^\circ\to\hbox{invisible})\ <\ 2.7\ 10^{-7}\ \ ,
\ee
at the 90 \% c.l.,
from the decay $K^+\to\pi^+\,+$ invisible $\,\pi^\circ$.
$\,U$ bosons that would be pair-produced in $\pi^\circ$ decays,
and remain undetected (as decaying into unobserved LDM or $\,\nu\bar\nu\,$ pairs), 
would have to verify

\vspace{-5mm}

\be
(f_{u}^2-f_{d}^2)^2\ <\ 2.7\ 10^{-7}\ \,\frac{e^4}{9}\ \simeq \ 2.5\ 10^{-10}\ \ ,
\ee
i.e

\vspace{-8mm}
\be
\label{limfufd}
\sqrt{|f_{u}^2-f_{d}^2|} \ \lsim\ 4\ 10^{-3}\ \ .
\ee

If the $U$ couples in the same way to the $u$ and $d$ quarks
the expected branching ratio for $\pi^\circ\to UU$ is very small, as follows
from the isospin 1 of the $\pi^\circ$ meson
(disregarding small isospin violations 
associated with $\,\pi^\circ$ mixing with $\eta$ or $\eta'$).
Except in the situations for which the $U$ couplings are close to isoscalar so that 
$\,f_u\simeq f_d$, then allowed to be significantly larger, we find the typical constraint
on the $U$ coupling to a quark
\be
\label{limfufd2}
|f_{q}|\ \lsim\ \ \,4\ 10^{-3}\ \ (\hbox{up to}\ \,\approx 10^{-2})\ \ .
\ee
which applies to both {\it vector} and {\it axial} couplings \,--  
provided of course the $U$ mass 
remains somewhat lighter than $\,m_{\pi^\circ}/2$.

\subsection{\boldmath $\,\pi^\circ \to \gamma \,U$}

We now turn to the mixed decay $\,\pi^\circ\to\gamma\,U$.
As $\,\pi^\circ$ and $\gamma$ have $C=+$ and $-$, respectively,
only the vectorial couplings of $U$ to quarks can now contribute to the amplitude.
We get, in a similar way,
\be
\ba{ccl}
\displaystyle
\frac{B(\pi^\circ\to \gamma \,U)}{B(\pi^\circ \to \gamma\,\gamma)}\ &\simeq&\ 
\displaystyle 2\ \left(\ \frac{(\frac{2e}{3} f_{uV}- \frac{-e}{3}f_{dV})}
{(2e/3)^2-(-e/3)^2}\ \right)^2
\vspace{4mm}\\
&\simeq&\ \displaystyle  \frac{18}{e^2}\ \ \left(\frac{2\,f_{uV}+f_{dV}}{3}\right)^2\ \ .
\ea
\ee
If the $U$ stays invisible, this decay rate should satisfy \cite{atyia},
\be
B(\pi^\circ\ \to\ \gamma +\ (\hbox{invisible}\ U))\ < \ 5\ 10^{-4}\ \ 
\ee
for a light $m_U$, the limit decreasing down to about 2\ $10^{-4}$ for $m_U \simeq 100$ MeV.
We then obtain, for the lighter $U$ masses, $\,[(2f_{uV}+f_{dV})/3]^2 < 2.5\ 10^{-6}$,
\,and therefore
\be
\label{limfufd3}
\frac{|2\,f_{uV}+f_{dV}|}{3} \ <\ 1.6\ 10^{-3}\ \ ,
\ee
valid as long as $m_U\lsim m_{\pi^\circ}/2$. 
This is one of the most restrictive bounds we get on vectorial quark couplings,
that we can remember (in a simplified way) as
\be
|f_{qV}| \ <\ 2\ 10^{-3}\ \ ,\ \ \ \ \hbox{or}\ \ \ \frac{f_{qV}^{\,2}}{4\pi}  \ <\ 2\ 10^{-7}\ \ .
\ee

\vspace{3mm}
If the $U$ were to decay into $e^+e^-$, these events would resemble to some extent 
(at least for the lighter $U$'s), events in which $\,\pi^\circ\to\gamma\,e^+e^-$
(which has a branching ratio of $(1.25\pm .04\pm.01)\ 10^{-2}$ \cite{gammaee}, 
in agreement with QED expectations).
This requires, for such light $U$ bosons,
\be
\label{limfufd4}
\frac{|2\,f_{uV}+f_{dV}|}{3} \ \lsim\ 5\ 10^{-3}\ \ ,
\ee
the limit being presumably more constraining for heavier $U$'s, as 
a significant production of $e^+e^-$ pairs with a larger invariant mass would probably have not stayed unnoticed.

\subsection{\boldmath $\,\eta\,$ ({\lowercase {or}} \boldmath $\,\eta'$) $\ \to \ UU\,$}

We can perform the same calculation for the isospin-0 state
$\,\eta= \cos\alpha\ \frac{u\bar u+d \bar d}{\sqrt2}-\sin\alpha\,s\bar s$
\,($\alpha$ including in particular $\eta$-$\eta'$ mixing effects). 
We can also extend it to $\eta'$, written as
$\,\eta'=  \cos\alpha'\,\left(\,\sin\alpha \ 
\frac{u\bar u+d \bar d}{\sqrt2}+\cos\alpha\,s\bar s\,\right)\ +\ \sin\alpha'\ (\,...\,)$ . 
We get
\be
\label{etadecay}
\ba{l}
\displaystyle
\frac{B(\eta\to \,U \,U)}{B(\eta \to \,\gamma\gamma)}\ 
\vspace{3mm}\\
\simeq\
\displaystyle \left(\ 
\frac{\frac{\cos\alpha}{\sqrt 2} \,(f_{u}^2+f_{d}^2)- \sin\alpha\, f_{s}^2 }
{\frac{\cos\alpha}{\sqrt 2}\,[(2e/3)^2+(-e/3)^2]-\sin\alpha\,(-e/3)^2}\ \right)^2 \ \ ,
\ea
\ee
and similarly for $\,\eta'$.
We shall suppose for simplicity that we are close to a situation of ideal mixing, 
with  $\alpha$ small, and $\eta$ not too far from $\frac{u\bar u+d \bar d}{\sqrt2}$.
Corrections may be taken into account immediately from (\ref{etadecay}), replacing $\,(f_u^{\,2}+f_d^{\,2})\,$ by the appropriate linear combination
with $\,f_s^{\,2}$\,. And similarly for $\eta'$.
This leads to the approximate expressions
\be
\ba{ccl}
\displaystyle
\frac{B(\eta\to \,U \,U)}{B(\eta \to \,\gamma\gamma)}\ 
&\simeq&\
\displaystyle 
\frac{(f_{u}^2+f_{d}^2)^2}
{\,[(2e/3)^2+(-e/3)^2\,]^2}
\vspace{4mm} \\
&\simeq&\
\displaystyle
\frac{81\ (f_{u}^2+f_{d}^2)^2}{25\ e^4}\ \ \ ,
\ea
\ee
and
\be
\ba{ccl}
\displaystyle
\frac{B(\eta'\to \,U \,U)}{B(\eta'\to \,\gamma\gamma)}\ 
&\simeq&\
\displaystyle \frac{81 \ f_{s}^4} {e^4}\ \ .
\ea
\ee

The recent experimental limits \cite{eta}
\be
\left\{\ 
\ba{ccl}
\displaystyle
\frac{B(\eta\to \,\hbox{invisible})}{B(\eta \to \,\gamma\gamma)}\ <\ 
1.65\ 10^{-3}\ \ 
\vspace{3mm}\\
\displaystyle
\frac{B(\eta'\to \,\hbox{invisible})}{B(\eta' \to \,\gamma\gamma)}\ <\ 
6.7\ 10^{-2}\ \ 
\ea \right.
\ee
imply directly, assuming that the $U$ boson remains invisible 
(as decaying for example into light dark matter particles):
\be
\label{limfufdfs}
\left\{\ 
\ba{ccl}
\sqrt{f_{u}^2+f_{d}^2}&< & 4.5\ 10^{-2}\ ,
\vspace{2mm}\\
|f_s| &< & 5\ 10^{-2}\ \ \ .
\ea\right.
\ee
These expressions have to be slightly adapted, as indicated earlier, 
so that the first formula constrains in fact the combination
$\,|\,f_{u}^2+f_{d}^2\,-\sqrt 2\,\tan\alpha \,f_s^{\,2}\,|$.
\,Again these limits apply to both {\it vector} and {\it axial} couplings.
They may look less constraining than the previous ones, but they also
apply to larger values of $m_U$.

\subsection{\boldmath $\pi^\circ,\ ...\ \to \ $ LDM particles \ (\,+\boldmath$\,\gamma$\,)}

We finally note that decays such as $\,\pi^\circ$ (or $\eta$, or $\eta'$) into a pair of LDM particles ($\chi\chi,\ \chi\bar\chi$, or $\varphi\bar\varphi$), without emitted photon, 
e.g.

\vspace{-6mm}

\be
\pi^\circ\ \to \chi\chi \ \ ,
\ee
can only proceed (owing to $C$) through an axial coupling of the $U$ to quarks, not a vector coupling. 
These decays will lead to limits for the products $|c_\chi\,f_{qA}|$, 
already strongly constrained as we have seen.
On the other hand decays such as
\be
\pi^\circ\ \to\ \gamma\ \,\chi\,\chi\ \ ,\ \ ...\ \ ,
\ee
can proceed through the vector coupling $\,f_{qV}$ (just as for $\,\pi^\circ\to\gamma\,U$), 
leading this time to limits on $|c_\chi\,f_{qV}|$.

\section{Conclusion}

Altogether the new limits (\ref{limcfc},\ref{limcfb},\ref{limfcv},\ref{limcfuv},\ref{limfuv},\ref{limfufd},\ref{limfufd2},\ref{limfufd3},\ref{limfufd4},\ref{limfufdfs}) 
involving the {\it vector} couplings of a $U$ to quarks
are at best of the order of a few $\,10^{-3}$, \,e.g. $1.6\ 10^{-3}$ in (\ref{limfufd3})
from $\pi^\circ\to\gamma\, +$ invisible $\,U$, for a $U$ sufficiently light compared to the $\pi^\circ$. 

\vspace{2mm}

Rare decays of quarkonia and mesons into invisible particles or $\gamma$ + invisible particles have provided, and will continue to provide, promising ways 
to search for new neutral ``weakly''-interacting light particles,
especially if the $U$ boson is not too light 
so that its couplings can more easily be larger, particularly the vectorial ones.
For the time being, the new constraints obtained on the vector couplings of quarks 
are considerably less restrictive than existing ones on {\it axial} 
couplings, from an axionlike behavior of a light $U$ boson, 
tested in $\,\psi\to\gamma \,U,\ \Upsilon\to \gamma \,U$ and $K^+\to\pi^+ U$
decays; or as compared to the limit from parity-violation in atomic physics,
in the presence of an axial coupling to the electron.
The new limits do not restrict significantly the 
properties of Light Dark Matter particles and $U$ bosons that would be responsible for their annihilations, or production;
especially as electron couplings (rather than quark couplings)
play a crucial role for LDM annihilations into $\,e^+e^-$. 
Altogether the hypothesis of a light neutral gauge boson $U$,
and light dark matter particles,
remains a fascinating possibility.


\begin{thebibliography}{}


\bibitem{susy} 
P.~Fayet, Phys. Lett. B {\bf 69}, 489 (1977); 
G.R.~Farrar and P.~Fayet, Phys. Lett. B {\bf 76}, 575 (1978); B {\bf 79}, 442 (1978).


\bibitem{grav} 
P.~Fayet, Phys. Lett. B {\bf 70}, 461 (1977); B {\bf86}, 272 (1979).



\bibitem{axion}
F.~Wilczek, Phys. Rev. Lett. {\bf 40}, 279 (1978); S.~Weinberg, Phys. Rev. Lett. {\bf 40}, 223 (1978).

\bibitem{fayet:1980rr}
P.~Fayet, Phys. Lett. B {\bf 95}, 285 (1980); Nucl. Phys. B {\bf 187}, 184 (1981).

\bibitem{boehmfayet}
  C.~Bo$\!$ehm and P.~Fayet, Nucl. Phys. B {\bf 683}, 219 (2004).

\bibitem{fermion}
  P.~Fayet,
  Phys.\ Rev.\ D {\bf 70}, 023514 (2004);
proc. 39th Renc. Moriond, hep-ph/0408357; hep-ph/0607094.


\bibitem{integral}
P. Jean {\em et al.}, Astron. Astrophys. {\bf 407}, L55 (2003).



\bibitem{boehm511}
  C.~Bo$\!$ehm {\em et al.}, 
  Phys.\ Rev.\ Lett.\  {\bf 92}, 101301 (2004);
  C.~Bo$\!$ehm, P.~Fayet and J.~Silk,
  Phys.\ Rev.\ D {\bf 69}, 101302 (2004).


\bibitem{cras}
P.~Fayet, C. R. Acad. Sci. t. 2, S\'erie IV, 1257 (2001), hep-ph/
0111282,
and refs. therein.



\bibitem{pvat}
P.~Fayet, Phys. Lett. B {\bf 96}, 83 (1980); 
C.~Bouchiat and C.A.~Piketty,  Phys. Lett. B {\bf 128}, 73 (1983);
C.~Bouchiat and P.~Fayet, Phys. Lett. B {\bf 608}, 87 (2005).




\bibitem{psi}
P.~Fayet, Phys. Lett. B {\bf 84}, 421 (1979).

\bibitem{breid}
SLAC-LBL coll., M. Breidenbach, priv. comm. (1979), unpub..


\bibitem{PDG}
Review of particle properties, Phys. Lett. B {\bf 204}, 262 (1988).


\bibitem{CLEO1}
D.~Besson {\em et al.}, CLEO Coll., Phys.\ Rev.\ D {\bf 30}, 1433 (1984) .

\bibitem{photinoprod}
P.~Fayet, Phys. Lett. B {\bf 117}, 460 (1982); B {\bf 175}, 471 (1986).


\bibitem{photinoprodexp}
E. Fernandez {\em et al.} Phys. Rev. Lett. {\bf 54}, 1118 (1985);
G. Bartha {\em et al.}, Phys. Rev. Lett. {\bf 56}, 685 (1986);
H.J.~Behrend {\em et al.}, CELLO Coll., Z. Phys. C {\bf 35}, 181 (1987);
C.~Hearty {\em et al.}, Phys. Rev. D {\bf 39}, 3207 (1989).




\bibitem{cosmion}
P.~Fayet and J. Kaplan, Phys. Lett. B {\bf 269}, 213 (1991).

\bibitem{elrath}
B.~McElrath, Phys. Rev. D {\bf 72}, 103508 (2005).



\bibitem{edwards}
C.~Edwards {\em et al.}, Phys.\ Rev.\ Lett.\  {\bf 48}, 903 (1982).

\bibitem{crystal}
D.~Antreasyan {\em et al.}, Crystal Ball Coll., 
{\em Phys. Lett.} B {\bf 251}, 204, 1990.


\bibitem{CLEO2}
R.~Balest {\em et al.}, CLEO Coll., Phys.\ Rev.\ D {\bf 51}, 2053 (1995).


\bibitem{rasera} Y.~Ascasibar {\em et al.},  astro-ph/0507142;
Y.~Rasera {\em et al.}, Phys.\ Rev.\ D {\bf 73}, 103518 (2006).

\bibitem{vector}
P.~Fayet, Phys. Lett. B {\bf 227}, 127 (1989);
Nucl. Phys. B {\bf 347}, 743 (1990).



\bibitem{kpiu}
Y.~Asano {\em et al.}, Phys.Lett. B {\bf 107}, 159 (1981);
S.~Adler \textit {et al.}, Phys. Rev. Lett. {\bf 88}, 041803 (2002); 
Phys. Lett. B {\bf 537}, 211 (2002);
V.V.~Anisimovsky \textit {et al.}, Phys. Rev. Lett. {\bf  93}, 031801 (2004).



\bibitem{artamonov}
A.~Artamonov {\em et al.}, Phys. Rev. D {\bf 72}, 091102 (2005).


\bibitem{atyia}
M.~Atiya {\em et al.}, Phys.\ Rev.\ Lett.\  {\bf 69}, 733 (1992).

\bibitem{gammaee}
M.A.~Schardt {\em et al.}, Phys.\ Rev.\ D {\bf 23}, 639 (1981).

\bibitem{eta}
M.~Ablikim {\em et al.}, BES Coll., hep-ex/0607006.




\end{thebibliography}
\end{document}